\documentclass[aps,prb amsmath,amssymb,11pt]{revtex4-1}

\usepackage{amsmath,amsfonts,amssymb}
\usepackage{graphicx}
\usepackage{subfigure}
\usepackage[colorlinks=true, allcolors=blue]{hyperref}
\usepackage[usenames, dvipsnames]{color}
\usepackage{siunitx}
\begin{document} 

\title{Tunable X-ray source by Thomson scattering during laser-wakefield acceleration}

\author{Sabine Schindler}
\address{Ludwig-Maximilians-Universit\"{a}t M\"{u}nchen, Am Coulombwall 1, 85748 Garching, Germany}
\address{Max Planck Institut f\"{u}r Quantenoptik, Hans-Kopfermann-Str. 1, 85748 Garching, Germany}
\author{Andreas D\"opp}
\address{Ludwig-Maximilians-Universit\"{a}t M\"{u}nchen, Am Coulombwall 1, 85748 Garching, Germany}
\address{Max Planck Institut f\"{u}r Quantenoptik, Hans-Kopfermann-Str. 1, 85748 Garching, Germany}
\author{Hao Ding}
\address{Ludwig-Maximilians-Universit\"{a}t M\"{u}nchen, Am Coulombwall 1, 85748 Garching, Germany}
\address{Max Planck Institut f\"{u}r Quantenoptik, Hans-Kopfermann-Str. 1, 85748 Garching, Germany}
\author{Max Gilljohann}
\address{Ludwig-Maximilians-Universit\"{a}t M\"{u}nchen, Am Coulombwall 1, 85748 Garching, Germany}
\address{Max Planck Institut f\"{u}r Quantenoptik, Hans-Kopfermann-Str. 1, 85748 Garching, Germany}
\author{Johannes G\"otzfried}
\address{Ludwig-Maximilians-Universit\"{a}t M\"{u}nchen, Am Coulombwall 1, 85748 Garching, Germany}
\address{Max Planck Institut f\"{u}r Quantenoptik, Hans-Kopfermann-Str. 1, 85748 Garching, Germany}
\author{Stefan Karsch}
\address{Ludwig-Maximilians-Universit\"{a}t M\"{u}nchen, Am Coulombwall 1, 85748 Garching, Germany}
\address{Max Planck Institut f\"{u}r Quantenoptik, Hans-Kopfermann-Str. 1, 85748 Garching, Germany}

\begin{abstract}
We report results on all-optical Thomson scattering intercepting the acceleration process in a laser wakefield accelerator. We show that the pulse collision position can be detected using transverse shadowgraphy which also facilitates alignment. As the electron beam energy is evolving inside the accelerator, the emitted spectrum changes with the scattering position. Such a configuration could be employed as accelerator diagnostic as well as reliable setup to generate x-rays with tunable energy. 

\end{abstract}

\maketitle

\section{INTRODUCTION}

The development of high-peak-power laser systems based on chirped-pulse amplification\cite{Strickland:1985th} has led to the emergence of a range of new technologies. Among these is laser wakefield acceleration (LWFA), a technology with the potential to provide compact and bright electron beams\cite{Malka:2008un,Hooker:2013jk}. Since the first demonstration of quasi-monoenergetic electron beams in 2004\cite{Faure:2004tj,Geddes:2004vs, Mangles:2004vr}, the performance of these accelerators has been continuously increasing and nowadays, LWFAs driven by petawatt lasers routinely reach multi-GeV energies\cite{Wang:2013el,Kim:2017il,Gonsalves:2019ht}. Recent developments also include the implementation of (combined) injection techniques\cite{Faure:2006vy,Buck:2013gs,Thaury:2015dq, Wenz:2019gc}, dephasing mitigation\cite{Guillaume:2015dia} and chirp compensation\cite{Dopp:2018kf}.

Furthermore, LWFA electrons gives rise to a number of secondary radiation sources. Transition radiation can be used to generate THz signals\cite{Leemans:2003ij}, 
which can among others be used to characterize the bunch duration of the laser-accelerated electrons\cite{Heigoldt:2015cd}. By sending few-hundred MeV electrons into a magnet undulator, synchrotron radiation in the ultraviolet regime\cite{Fuchs:2009da,Anania:2014ew} can be generated. Potentially, this technology can also be extended to generate bright, coherent radiation in a table-top free-electron laser \cite{Couprie:2015te}. In the x-ray and $\gamma$-ray regime, three different methods exist to generate high energy photon beams, namely Bremsstrahlung conversion\cite{Glinec:2005ve,Dopp:2016faa}, Betatron radiation\cite{Rousse:2004tc,Dopp:2017dza} and Thomson scattering\cite{Schwoerer:2006dw,TaPhuoc:2012cg,Powers:2013bx,Khrennikov:2015gxa,Sarri_2014}. Betatron radiation is wiggler-like radiation that occurs as the result of transverse oscillations of electrons inside the laser wakefield. With a photon yield of up to $10^9$ photons per shot and a micrometer source size, this radiation can readily be used for propagation-based phase contrast imaging\cite{Fourmaux:2011cs,Kneip:2011cx}. The technique has already been extended to perform phase-contrast tomographies\cite{Wenz:2015if} and in recent experiments the acquisition time for tomographic data sets has been reduced to a few minutes\cite{Dopp:2018ub,Gotzfried:2018cx}. However, betatron radiation is not suitable for advanced imaging techniques such as dual-energy imaging, because of its large bandwidth.

Thomson sources, sometimes also referred to as inverse Compton sources, potentially offer greater flexibility in this regard. While initially studied at conventional accelerator facilities, in particular in order to generate femtosecond radiation\cite{Schoenlein:1996vz}, recent years have seen an increase of studies on all-optical Thomson sources. These use one laser for wakefield acceleration and either the same\cite{TaPhuoc:2012cg} or an additional laser pulse\cite{Khrennikov:2015gxa,Powers:2013bx} for scattering, allowing for a very compact and flexible setup.

Thomson scattering provides an elegant way to upshift the frequency of laser light. Electrons perceive the laser frequency $\lambda_0$ doppler-upshifted by a factor of $\gamma$ and emit radiation at the wavelength $\lambda / \gamma$, where $\gamma$ denotes the Lorentz factor. An observer at rest receives this radiation upshifted by another factor of $\gamma$. Furthermore, the measured wavelength depends on the angle between electrons and laser $\phi$, the angle to the observer $\theta$ and the normalized vector potential $a_0$. The fundamental wavelength of the Thomson scattered radiation therefore is \cite{Ride:1995vh}

\begin{equation}
\label{eqn:thomson}
    \lambda_{TS} = \frac{\lambda_0}{2 \gamma^2 (1 - \beta \cos \phi)} \left(1+\frac{a_0^2}{2} + \gamma^2 \theta^2 \right)
\end{equation}

\begin{figure} [t]
    \begin{center}
    \includegraphics[width=0.99\textwidth]{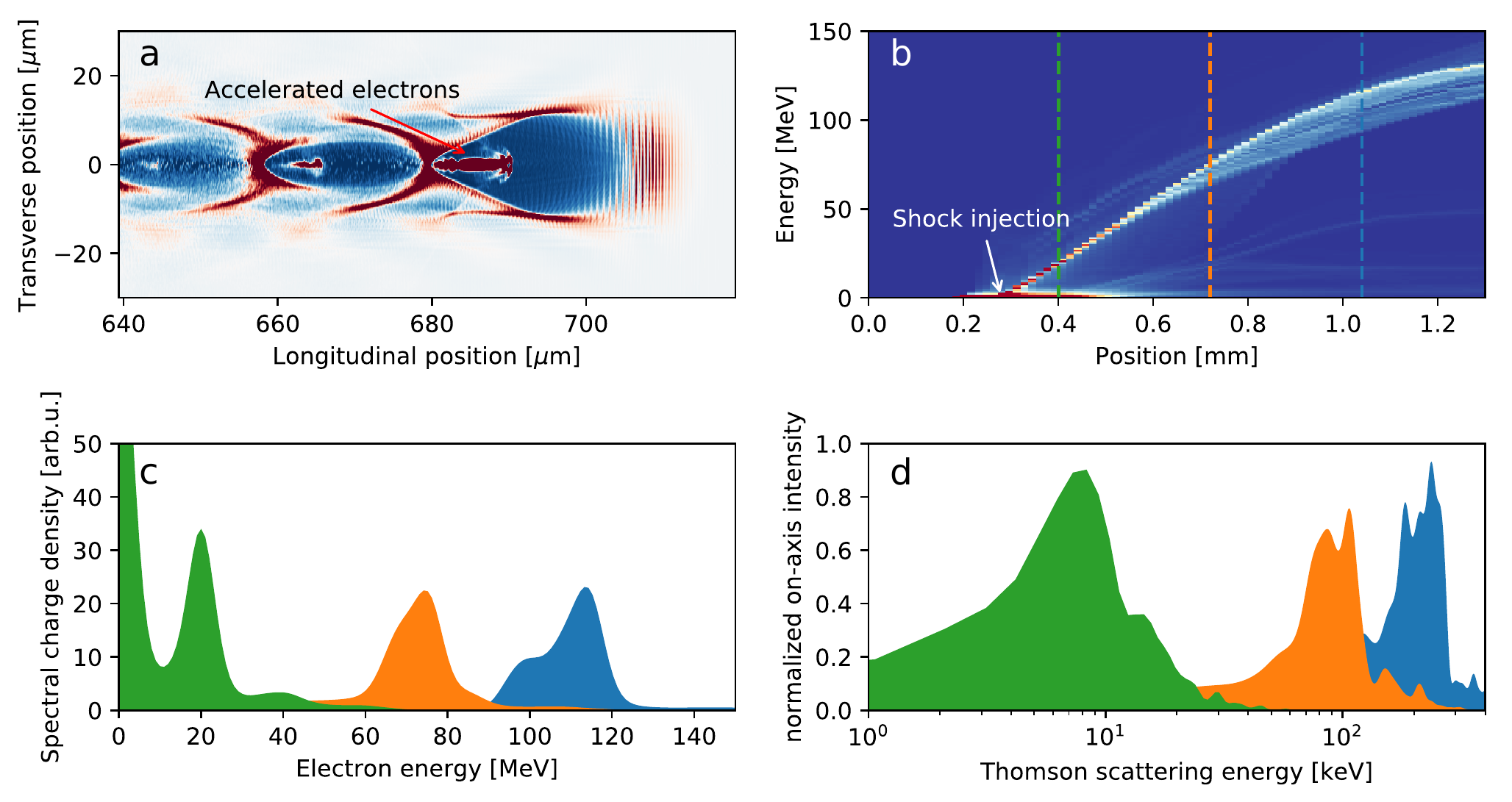}
    \end{center}
    \caption[example] 
    {\label{fig:simulations} 
    Particle-in-cell simulations of a laser wakefield accelerator employing shock-front injection. (a): Snapshot of the electron density distribution after 0.7 mm of propagation. (b): Evolution of the electron energy spectrum. (c)+(d): Electron and resulting Thomson spectra for three different positions along the accelerator, corresponding to the dashed lines in (b).}
\end{figure}

In the following we present preliminary results on Thomson scattering not with an LWFA-accelerated electron bunch in free space, but with a bunch during acceleration in the LWFA itself. This configuration differs from usual experiments because the electron beam energy is still evolving inside the accelerator. To visualize the concept, we have simulated a laser wakefield accelerator with shock-front injector using the quasi-3D particle-in-cell code FBPIC\cite{Lehe:2016dn}. A snapshot of the electron density distribution and the evolution of the electron energy spectrum are shown in Fig. \ref{fig:simulations} on the top row. Electrons are injected at $z=\SI{0.25}{\milli\meter}$ and rapidly gain energy until they start to dephase. The energy of the quasi-monoenergetic peak changes from $E_{\SI{0.4}{\milli\meter}}=\SI{20}{\mega\electronvolt}$ to $E_{\SI{1.0}{\milli\meter}}>\SI{100}{\mega\electronvolt}$. Accordingly, the energy of the backscattered photons will increase by a factor of $(E_{\SI{1.0}{\milli\meter}}/E_{\SI{0.4}{\milli\meter}})^2>25$ (see Fig. \ref{fig:simulations} bottom row). We have calculated the backscattering signal using the electron beam parameters at different positions along the simulation and a scattering pulse with $a_0=0.5$ using the code \textsc{Chimera} \cite{Andriyash:2015th}. These simulations also show the wide tuning range of such a Thomson source. The geometry therefore has the potential to tune the energy of Thomson backscattering sources over a wide energy range, while maintaining the laser wakefield accelerator at stable, unmodified conditions. Furthermore, if the x-ray spectrum can be accurately measured the method could be applied as an accelerator diagnostic to measure the electron energy evolution inside the accelerator.

\section{Experimental Setup}

The experiments were performed using the ATLAS Ti:Sapphire laser system, situated at the Laboratory for Extreme Photonics (LEX) at Ludwig-Maximilians University of Munich (LMU), which delivered up to \SI{2.5}{\joule} pulses within \SI{28}{\femto\second} duration to the target chamber, resulting in a peak power of \SI{80}{\tera\watt}. In the experimental chamber a part of the beam, containing $\sim$ \SI{0.3}{\joule}, was split off by a pick-off mirror to provide a second head-on collision pulse. The driver pulse for electron acceleration was focused with a F/25 off-axis parabola to $a_0 \simeq 1.6$ on a $\mathrm{H}_2$ gas jet from a \SI{1}{\milli\meter} supersonic de Laval nozzle with an adjustable silicon wafer for shock-front injection and a plasma density of $n_0 = 4.5 \times 10^{18} \mathrm{cm}^{-3}$. The colliding pulse was focused by an F/26 off-axis parabolic mirror into the gas jet behind the shock-front to $a_0 \simeq 0.8$. The parabola was mounted on a translation stage to adjust the collision positions of the undulator beam with the accelerating electrons. 

The beam overlap in transverse direction was monitored with a top-view camera and shadowgraphy was used to control the temporal overlap hence the collision point of the beams.

As an electron diagnostic a \SI{80}{\centi\meter} permanent magnet (\SI{0.85}{\tesla}) in combination with a calibrated scintillating screen was used, measuring the electron bunch spectrum and charge content above the low-energy cut-off of \SI{27}{\mega\electronvolt}. 

The Thomson-scattered x-ray radiation was detected using a $\mathrm{Gd_2 O_2 S{:}Tb}$ based scintillator (P43), which is fiber-coupled to a 1388 $\times$ 1038 pixel CCD camera. The camera uses a $50:11$ taper, where each pixel represents an area of \SI{29}{\micro\meter\squared} on the scintillator screen. Placed at \SI{435}{\centi\meter} from the source, the field of view covers approximately 9.2 $\times$ \SI{6.9}{\milli\radian\squared}. At the end of the target chamber, \SI{30}{\centi\meter} in front of the camera, an aluminium filter-cake was placed in a attempt to determine the x-ray energy (shown in Fig. \ref{fig:Xray_images}).  

\vspace{0.5 cm} 

\begin{figure} [t]
    \begin{center}
    \includegraphics[width=0.99\textwidth]{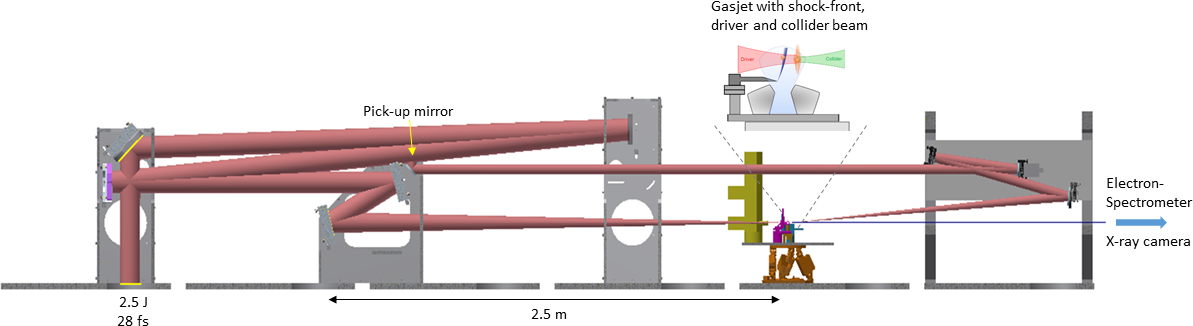}
    \end{center}
    \caption[example] 
    {\label{fig:setup} 
    Experimental setup: The main beam get split into a drive beam for LWFA and a head-on collision beam for Thomson scattering. The inset shows the target with the shock-front and the two colliding beams.}
    
\end{figure}

\section{Experimental Results}

We performed a series of 360 consecutive shots at 18 different collision positions (20 shots per position), step-wise increasing the distance between the electron injection point at the shock-front and the interaction point with the colliding laser beam. For each shot we recorded the x-ray signal emitted by the electrons at each respective stage of the acceleration process. The distance between the shock-front and the collision position was tuned from 200 to \SI{720}{\micro\meter} in steps of \SI{30}{\micro\meter}.

\subsection{Shadowgraphy imaging}\label{sec:probe}

We use shadowgraphy as a diagnostic for the plasma accelerator. In the shadowgrams we observe two distinctive diffraction features. The first from the left is the diffraction caused by the supersonic shock in the nozzle, i.e. diffraction of the probe beam at the transition from high plasma density to lower plasma density. The angle with respect to the propagation axis depends on the position of the blade inside the jet. At the position of the shock we also observe a bright light emission. This signal is associated with the electron injection process, known as wave-breaking radiation \cite{Thomas:2007cc}.

\begin{figure} [t]
    \begin{center}
    \subfigure{\includegraphics[width=0.24\textwidth]{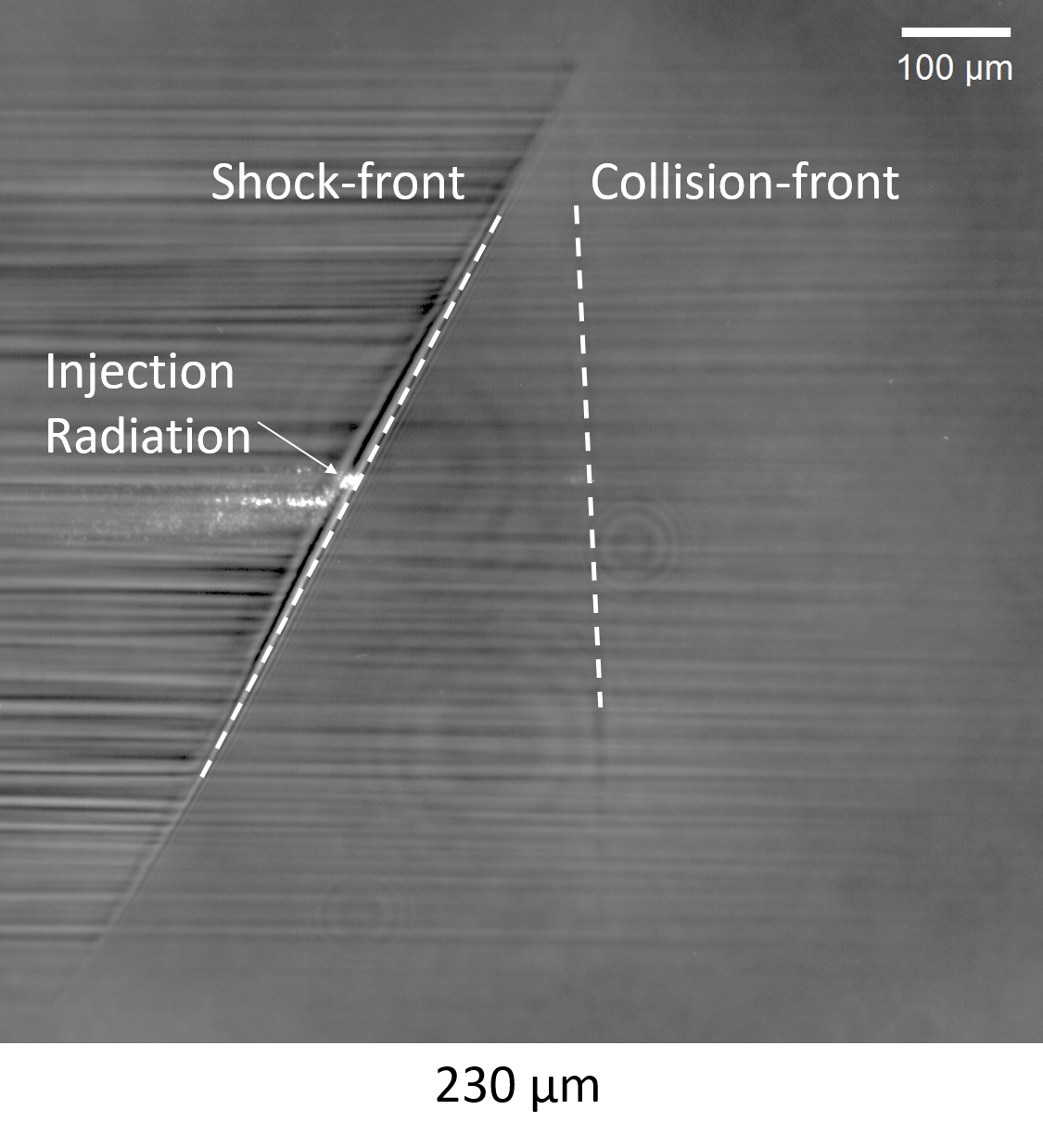}}
    \subfigure{\includegraphics[width=0.24\textwidth]{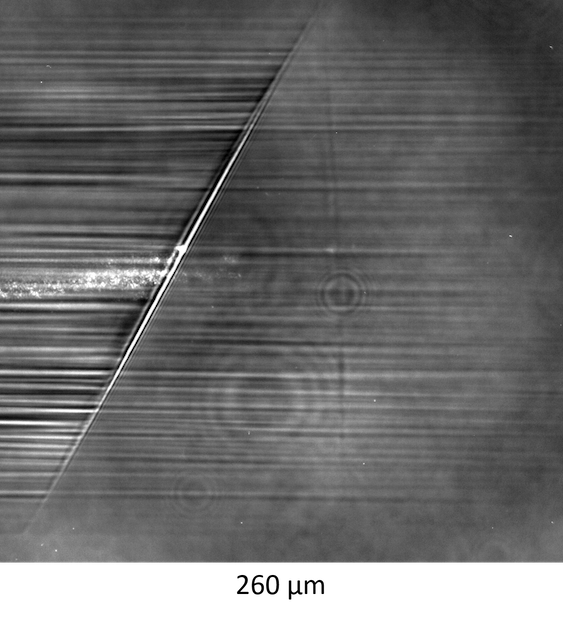}}
    \subfigure{\includegraphics[width=0.24\textwidth]{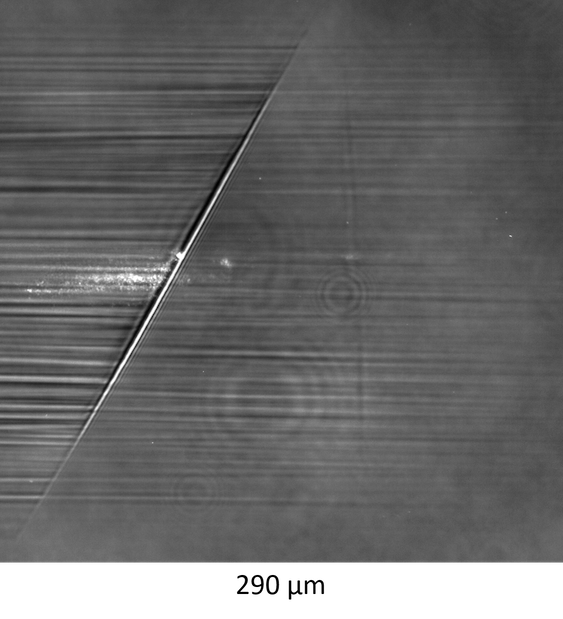}}
    \subfigure{\includegraphics[width=0.24\textwidth]{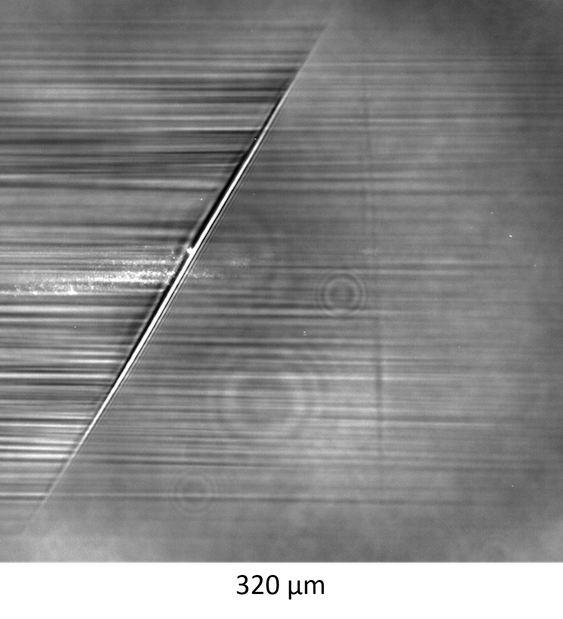}}
    \end{center}
    \caption[example] 
    {\label{fig:probe} 
    Shadowgraphy images of four different collision positions at distances from 230 to \SI{320}{\micro\meter} behind the shock-front: The electron driving beam traveling from left to right with the wafer generated shock-front where injection radiation was observed, right of it the collision feature of the two opposing beams is visible, showing the location of the radiation generation. The collision-front at the different positions is clearly moving to the right as the shock-front is fixed (note the relative position to the circular diffraction rings in the lower half of the image).}
\end{figure}

A second, weaker diffraction feature is produced once the colliding pulse is activated. The feature moves along the propagation as the timing of the collider is adjusted and marks the collision point of the two beams. This previously unobserved phenomenon allows us to accurately determine the distance between the shock, i.e. electron injection, and the pulse collision, i.e. the scattering position. Shadowgraphy images of four consecutive positions are shown in Fig. \ref{fig:probe}.

The formation of the second diffraction feature is likely related to the ionization dynamics of hydrogen. First it is important to note that the region of ionization (as indicated by the horizontal striations) extends well beyond the diameter of the main laser mode, because even weak spatial wings of the pulse are sufficient to partially ionize hydrogen gas. In contrast to the simple case of atomic hydrogen (with the well-known binding energy of \SI{13.6}{\electronvolt}), the ionization of hydrogen molecules is more complex. After ionization of one electron, the remaining $H_2^+$ molecule has a ground state energy of \SI{15.4}{\electronvolt} \cite{Sharp:1970kg} and hence requires higher intensities for multi-photon or tunneling ionization. We therefore expect that the plasma away from the laser axis consists mainly of $H_2^+$ (which later dissociates into $H^+$ and $H$). As the pulse collision produces locally increased electric fields, it is expected that hydrogen can be completely ionized. This will result in a local electron density peak, which will diffract the probe beam. This conclusion is also supported by the fact that the second diffraction feature gets fainter close to the laser axis where the intensity of the wings is higher. Therefore, the main driver will cause a higher degree of ionization, an the relative enhancement of the ionization degree by the colliding pulse becomes less pronounced. However, a more detailed study will be necessary to verify this interpretation of the results.

\subsection{Electrons}

\begin{figure} [ht]
    \begin{center}
    \includegraphics[height=6cm]{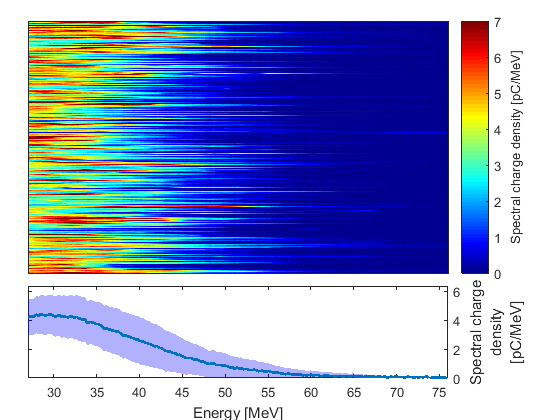}
    \end{center}
    \caption[example] 
    {\label{fig:electrons} 
    Top: Measured electron spectra of 360 consecutive shots. Bottom: Average electron spectrum of all shots, the shaded area indicates the standard deviation.}
\end{figure}

We generated shock-front induced electrons with a mean energy of \SI{30}{\mega\electronvolt}. The 360 single shot spectra and the mean and standard deviation of the average total spectrum are shown in Fig. \ref{fig:electrons}. The mean charge of the electrons hitting the scintillator was \SI{55}{\pico\coulomb}. As the accelerator was operated at the low energy limit to keep the scattered x-rays in a suitable energy range for the filter and detector, the stability, divergence and bandwidth of the generated electrons was not optimal.

Because we used a small gas jet with a total length well below the dephasing limit, the electrons constantly gain energy in the gas jet. Therefore we can estimate an upper energy limit of the generated photons using the final measured electron spectra and equation \ref{eqn:thomson}. With a laser potential of $a_0 = 0.8$ for the colliding beam like in our experiment, scattered x-rays up to \SI{46}{\kilo\electronvolt} can be achieved with the \SI{50}{\mega\electronvolt} electrons at the end of the accelerator. The maximum energy for the electron energy peak with \SI{30}{\mega\electronvolt} would be \SI{17}{\kilo\electronvolt}. As we scatter during the acceleration process, and hence with lower energetic electrons, we expect the measured x-ray energy to be much lower than this upper limit. However this estimate only takes into account photons at the fundamental frequency, while higher harmonics cause a broadening of the spectrum and a shift to higher energies. At the collider intensity used in the experiment, we expect substantial emission also at the 2nd (due to finite beam divergence) and 3rd harmonics, leading to a further broadening of the spectrum towards high energies. With such broadband spectra to start with, it is therefore difficult to distinguish photon spectra for different electron energies.   

\subsection{X-rays}

\begin{figure} [ht]
    \begin{center}
    \subfigure{\includegraphics[width=0.49\textwidth]{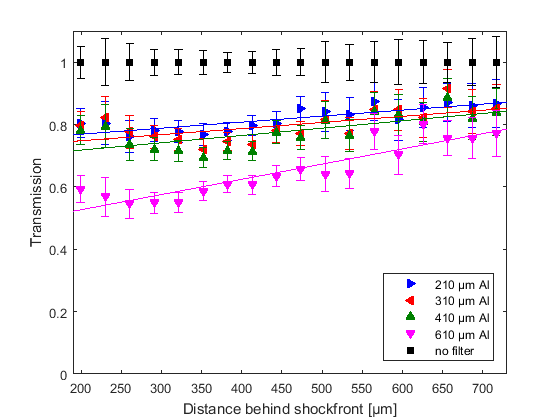}}
    \subfigure{\includegraphics[width=0.49\textwidth]{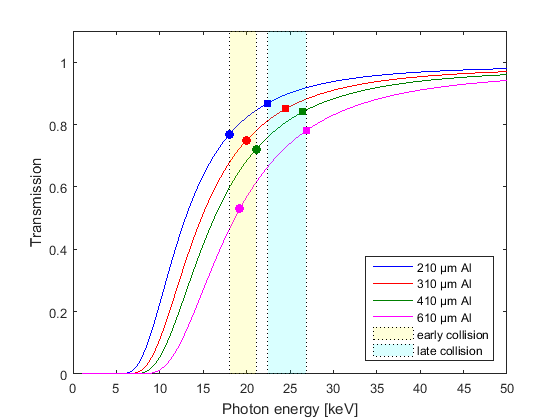}}
    \end{center}
    \caption[example] 
    {\label{fig:Filter_cts} 
    Left: Mean relative transmission of the four thickest filters for different collision positions and the corresponding linear fits. The transmission of the each filter clearly increases with increasing distance of the collision point to the shock-front, as expected from the electron energy gain in the progressive acceleration process. Right: Radiation transmission curves of aluminum with the thicknesses of the four analyzed filter parts \cite{NIST-Abs-coeff}. The circles and squares indicate the filter transmission values of the linear fits at the first and last collision position respectively.}
    
\end{figure}

The filter-cake for the estimation of the photon energy of the generated x-rays consists of 14 aluminium sectors with thicknesses of 5 to \SI{610}{\micro\meter}, which are suitable to attenuate photon energies in the range of 2 to \SI{40}{\kilo\electronvolt}. Additionally, a \SI{4}{\milli\meter} copper sector was also used to provide discrimination against bremsstrahlung photons. For the study of the spectral evolution we analyzed only the four most absorbing aluminium parts with thicknesses of 210 to \SI{610}{\micro\meter} as the thinner sectors are highly transparent in all cases and cannot provide clear results in the transmission trend. The radiation transmission curves of the analyzed filter sectors are displayed on the right side of Fig. \ref{fig:Filter_cts} for x-ray energies ranging from 1 to \SI{50}{\kilo\electronvolt}. 

\begin{figure} [ht]
    \begin{center}
    \includegraphics[width=0.99\textwidth]{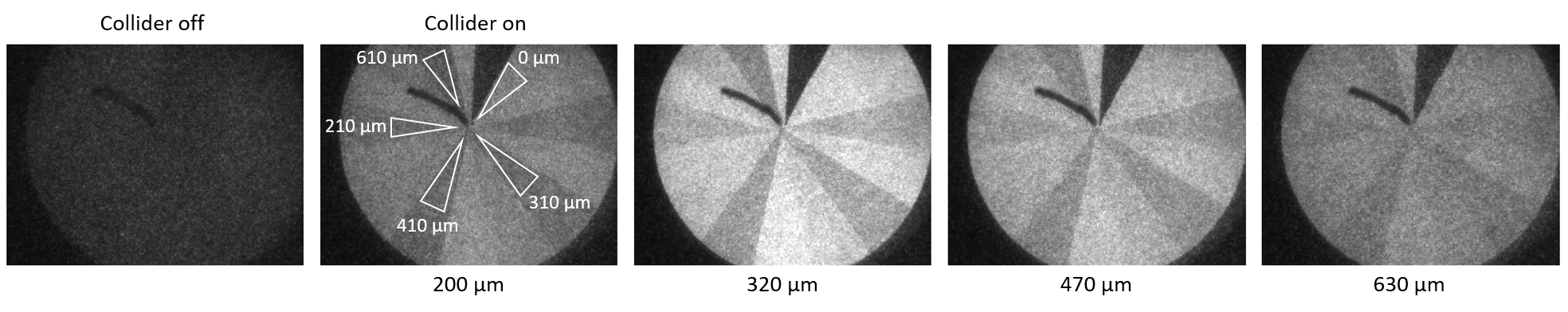}
    \end{center}   
    \caption[example] 
    {\label{fig:Xray_images} X-ray camera images for different Thomson scattering settings. The curved dark region in all images is a damaged spot on the scintillator, the dark triangle sticking in from the top is a \SI{4}{\milli\meter} thick copper sector.  The field of view of the camera covers a solid angle of 9.2 $\times$ \SI{6.9}{\milli\radian\squared}.
    Left: Average background image with blocked colliding beam shows only high-energy bremsstrahlung produced by electrons hitting the spectrometer magnet. Note that the \SI{4}{\milli\meter} copper piece is almost completely transparent. Right: X-ray average images of 20 shots each of the aluminum filter-cake at beam collision positions 200, 320, 470 and \SI{630}{\micro\meter} behind the shock-front.  }
\end{figure}

The average images of 20 shots at four different collision positions as well as the average reference image without the colliding beam are shown in Fig. \ref{fig:Xray_images}. The analyzed filter areas are highlighted with the corresponding thicknesses. 

We observe a noticeable bremsstrahlung background due to electrons hitting the spectrometer magnet because of their high divergence. When the colliding pulse was switched on, the overall transmission through the filter cake first increases when the collision point moves further downstream of the injection point due to the increasing photon energy, but then the signal falls off again due to the lower sensitivity of the camera for higher energy photons. However, the contrast between the different filters decreases as well, indicating a hardening of the photon spectrum.

For the x-ray analysis the weighted mean of the transmitted signal level in each filter part relative to the blank sector (\SI{0}{\micro\meter} at approx 1 o'clock in Fig. \ref{fig:Xray_images}) was calculated for 20 shots per collision position. The error bars indicate the fluctuation of the background bremsstrahlung which was assumed to be the highest uncertainty factor in this experiment as the signal level was low because of the large distance between x-ray source and detector. The results for all filters and collision positions are shown on the left side of Fig. \ref{fig:Filter_cts}. 

The relative transmission of all analyzed filter sections increases with increasing distance to the shock-front as a consequence of the electron energy gain in the progressive acceleration process. Deviations of the mean transmission at different collision positions are caused by electron and pointing fluctuations. 
For each filter, the measured transmission value given by the linear fit at \SI{200}{\micro\meter} and \SI{720}{\micro\meter} are indicated on the transmission curves in the right hand plot as circles and squares, respectively. While each filter transmission value gives a slightly different energy reading on the horizontal axis, all values shift by roughly the same amount, namely \SIrange{4.5}{7.5}{\kilo\electronvolt}. While due to the large photon energy spread these numbers cannot serve as a quantitative measure, they do indicate the expected hardening of the radiation spectrum with increasing accelerator length.   



\section{Conclusion \& Outlook}

In this experiment we demonstrated that Thomson scattering in the gas target during LWFA can be used as a tunable x-ray source. We were able to tune and monitor the collision point of the electrons with the colliding beam and observed an increase of the photon energy with increasing distance between the injection and collision points, as expected. Unfortunately the filter-cake was not optimal for adequate reconstruction of the x-ray spectrum in this photon energy range. With a more accurate spectral diagnostic the same technique becomes a potential tool to determine the electron energy at different stages during the acceleration process.   

At the Centre for Advanced Laser Applications (CALA) with the new ATLAS 3000 laser system we plan to further improve this setup to use it as an electron diagnostic and tunable x-ray source. With the higher laser power we could achieve electron beams with more charge resulting in a higher photon flux on the detector and more precise data. This necessitates the development of a more appropriate filter stack or alternative spectroscopy methods in order to better quantify the x-ray spectrum.

\appendix

\acknowledgments  
 
This work was supported by DFG through the Cluster of Excellence Munich-Centre for Advanced Photonics (MAP EXC 158) and SFB-Transregio TR-18 funding schemes, by EURATOM-IPP and the Max-Planck-Society. 


%

\end{document}